\documentclass[12pt]{iopart}

\usepackage{amssymb}
\usepackage{graphicx}
\begin{document}

\title{Quasienergy spectrum and tunneling current in ac-driven triple quantum dot shuttles}

\author{J Villavicencio$^1$, I Maldonado$^2$, E Cota$^3$ and G Platero$^4$  }
\address{$^1$ Facultad de Ciencias, Universidad Aut\'onoma de Baja
California, Ensenada, M\'exico}
\ead{villavics@uabc.edu.mx}
%
%
\address{$^2$ Centro de Investigaci\'on Cient\'ifica y de Educaci\'on
Superior de Ensenada, M\'exico}
\address{$^3$ Centro de Nanociencias y Nanotecnolog\'ia, Universidad
Nacional Aut\'onoma de M\'exico, Ensenada, M\'exico}
\address{$^4$ Instituto de Ciencia de Materiales de Madrid (CSIC),
Cantoblanco, 28049 Madrid, Spain}

\date{\today}
\begin{abstract}
The dynamics of electrons in ac driven double quantum dots have been extensively analyzed by means of  Floquet theory. In these systems, coherent destruction of tunneling has been shown to occur for certain ac field parameters. In the present work we analyze, by means of Floquet theory, the electron dynamics of a triple quantum dot in series attached to electric contacts, where the central dot position oscillates. In particular, we analyze the quasienergy spectrum of this ac driven nanoelectromechanical system, as a function of the intensity and frequency of the ac field and of external dc voltages.
For strong driving fields, we derive, by means of
perturbation theory, analytical expressions for the quasienergies
of the driven oscillator system.
From this analysis we discuss  the conditions for coherent
destruction of tunneling (CDT) to occur as a function of detuning and field parameters.
For zero detuning, and from the invariance of the Floquet Hamiltonian under a generalized parity transformation, we
find analytical expressions describing the symmetry properties of
the Fourier components of the Floquet states under such
transformation. By using these expressions, we show that in the
vicinity of the CDT condition, the quasienergy spectrum exhibits
exact crossings which can be characterized by the parity properties
of the corresponding eigenvectors.
\end{abstract}
%
\submitto{\NJP}
\maketitle

\section{Introduction}

Triple quantum dot systems have been realized in the laboratory and their stability diagrams \cite {gaudreau}, charge \cite{Rogge, kostyrko, fernando} and spin \cite{hawrylak, maria} transport properties   studied as a function of external parameters. These systems are important as components of quantum information circuits and molecular electronics. When one of the dots is set in motion by means of an external interaction, the system becomes a nanoelectromechanical system (NEMS). The study of these systems
has become an important area of research,
with potential applications as ultrasensitive detectors of
mass \cite{ekinci}, force \cite{rugar}, and displacement \cite{knobel}.
Recently, Azuma \etal \cite{azuma} have measured
quantized tunneling current from an $Au$ nanodot mounted on a
cantilever, using a scanning tunneling microscope probe. Also,
Koenig \etal \cite{koenig} reported the fabrication of a mechanical
single electron transistor using ultrasound waves as the excitation
mechanism. In all these works, a
central role is played by an oscillating element acting as a shuttle
carrying one electron from one contact to the other. The first
proposal of such a device was made by Gorelik \etal \cite{gorelik} to
be used as a single electron transistor.

Ac driven quantum dot arrays have been extensively studied in recent years both experimentally \cite{kierig,dellavalle,lignier} and theoretically \cite{cref_plat,theoretical2W}.  An important physical effect analyzed in these systems is the electron localization induced by an ac field, where for special parameter values of the driving field corresponding to roots of the Bessel functions \cite{note1}, tunneling is suppressed. This effect, termed coherent destruction of tunneling, has been analyzed by means of Floquet theory and the corresponding quasienergy spectrum \cite{grossman}. Close approaches of the Floquet quasienergies, as the system parameters are varied, produce large modifications of the tunneling rate and hence of the dynamics of the system. In particular, when two quasi-energies are close to degeneracy, the time-scale for tunneling between the states becomes extremely long, producing the phenomenon of CDT \cite{noteCDT}.

In the present work, we extend the Floquet analysis \cite{cref_plat, grifoni,holthaus,fromherz,hanggi98} to ac driven vibrating quantum dots arrays. In particular, we analyze the quasienergy spectrum of a Triple Dot Quantum Shuttle (TDQS), consisting of three quantum dots in series, with one level each, called $\varepsilon_l, \varepsilon_c$,  and $\varepsilon_r$, respectively. The central dot oscillates with frequency $\omega$, and the system is in the presence of an ac-field characterized by frequency $\omega_{ac}$ and intensity $V_{ac}$ (figure \ref{Fig1}).
Additionally, there is an infinite bias applied to the contacts of the triple quantum dot system, and there can be a voltage difference (detuning) $\varepsilon_b=\varepsilon_l-\varepsilon_r$ between the left and right quantum dots. We refer to this setup as the \emph{dynamic case}, to distinguish it from the \emph{static case} where the central dot is fixed.
For {\it static} three or more quantum dots in series, detailed studies \cite{crefPRB2004,longhi,villas-boas} have been carried out looking for possible control of the electronic current features of the system from the analysis of crossings and anticrossings in the quasienergy spectrum.
Here we study, both
analytically, using first order perturbation theory,  and
numerically, the conditions for CDT in a TDQS, in terms of the
quasienergies as a function of detuning, $\varepsilon_b$, and
the driving field parameters, in {\it both} static ($\omega=0$) and
dynamic ($\omega\neq 0$) cases. For finite detuning, in the static case,
we obtain numerical results in agreement with a first order
perturbative calculation in the weak tunneling regime, and find
conditions where exact crossings of the quasienergies take place
corresponding to CDT.
In the dynamic case, we obtain different
results depending on whether the current resonance
as a function of detuning $\varepsilon_b$  is calculated at
even or odd multiples of the driving frequency $\omega_{ac}$.
For the former case,
anticrossings are obtained at values of the driving field parameters
corresponding approximately, as we will see below, to a root of a
Bessel function. We will show how, in the case of weakly coupled
quantum dots, these results can be reproduced with great accuracy
considering interdot tunneling in first order perturbation theory.
The resonances at odd multiples of $\omega_{ac}$ are shown to depend strongly on interdot tunnel coupling.
For zero detuning, we note that the Floquet Hamiltonian is symmetric under a Generalized Parity Transformation (GPT), and
the CDT phenomena can be explained by analyzing the quasienergy spectrum and the
parity of the corresponding Fourier coefficients of the Floquet states.

In section \ref{model} we define the formalism used to describe our system with special emphasis in Floquet theory. We present our results in section \ref{results}, and finally the conclusions in section \ref{conclusions}.


\section{\label{model}Model}

\begin{figure}[!tbp]
\center
\rotatebox{0}{\includegraphics[width=3.5in]{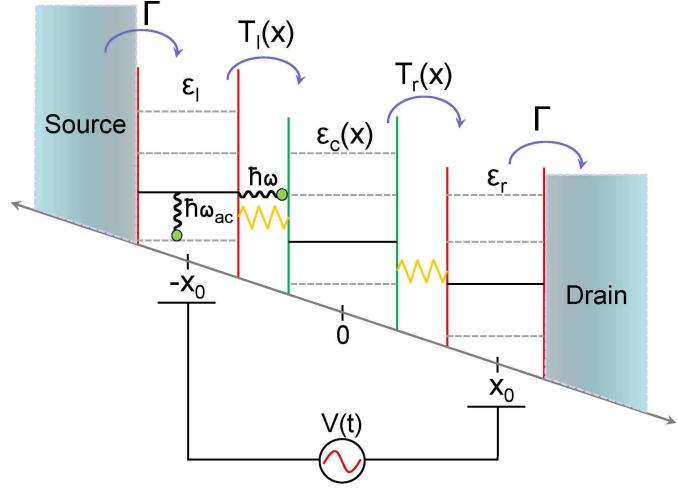}}
\caption{ Driven TDQS where oscillating tunneling barriers
between the central quantum dot and its neighbors are considered.
The system is driven by an ac-potential $V(t)=V_{ac}\,
\cos(\omega_{ac} t)$, and the left and right tunneling barriers are
rigidly coupled to the leads with constant transition rates, $\Gamma$.
The system is in an infinite bias setup, and there can be a voltage difference (detuning), $\varepsilon_b=\varepsilon_l-\varepsilon_r$, between left and right quantum dots.
 }
\label{Fig1}
\end{figure}

We consider a linear array of three quantum dots where the central movable dot is flanked by two static
dots at fixed positions $\pm x_0$ (figure \ref{Fig1}). The
oscillation of the central dot affects the tunneling rates,
which are now position dependent.
The charging energy and interdot Coulomb interaction are assumed to be sufficiently large that only
one transport electron can occupy the chain of three dots at any
given time (Coulomb blockade regime).
The TDQS is modelled by means of a
tight-binding
Hamiltonian $\hat{H}_{TQS}=(\hat{H}_0+\hat{H}_{osc}+\hat{H}_{tun})$
where
\begin{eqnarray}
\hat{H}_{0} &=&\varepsilon _{l}\left| l\right\rangle \left\langle
l\right| +\varepsilon _{c}(\hat{x})\left| c\right\rangle
\left\langle c\right|
+\varepsilon _{r}\left| r\right\rangle \left\langle r\right| ; \nonumber \\
\hat{H}_{osc} &=&\hbar \omega \,d^{\dagger }d; \nonumber \\
\hat{H}_{tun} &=&T_{l}(\hat{x})\left[ \left| l\right\rangle
\left\langle c\right| +\left| c\right\rangle \left\langle l\right|
\right] + \\ \nonumber &&T_{r}(\hat{x})\left[ \left| c\right\rangle
\left\langle r\right| +\left| r\right\rangle \left\langle c\right|
\right]. \label{hams am}
\end{eqnarray}

In the above equation, $\hat{H}_{0}$ represents the Hamiltonian
of the three dots. Here the energy of the central dot $\varepsilon
_{c}$ depends on the position operator $\hat{x}$ given by $\hat{x}=\Delta
x_{zp}\,(\hat{d}^{\dag}+\hat{d})$, where $\Delta
x_{zp}=(\hbar/2m\omega)^{1/2}$ is the zero-point uncertainty
position of the oscillator. Thus
$\varepsilon_c(\hat{x})=[\varepsilon_l-(\varepsilon_b/2x_0)(
\hat{x}+x_0)]$,
where it is assumed it undergoes a Stark shift proportional to its
position due to the detuning across the device, $\varepsilon_b=(\varepsilon_l-\varepsilon_r)$;
$\hat{H}_{osc}$ is the Hamiltonian associated to the oscillation of the central quantum dot (frequency $\omega$) where $d^{\dagger}, d$ are creation and destruction boson operators respectively; $\hat{H}_{tun}$ represents the tunneling between all three dots, with position dependent tunneling rates $T_r(\hat{x})$, and $T_l(\hat{x})$,  given by
\begin{eqnarray}
T_l(\hat{x})&=&-{\cal V} e^{-\alpha(x_0+\hat{x})}; \nonumber \\
T_r(\hat{x})&=&-{\cal V} e^{-\alpha(x_0-\hat{x})}, \label{tdefinition}
\end{eqnarray}
where ${\cal V}$ is the tunneling amplitude, and $\alpha$ is the inverse of
the tunneling length.
The position operator $\hat{x}$ measures the displacement of the
vibrational mode.
The matrix elements of the tunneling couplings $T^{ij}_l$ and $T^{ij}_r$ in the oscillator basis
can be calculated analytically by using a procedure discussed in reference \cite{osc1} which yields,
\begin{eqnarray}
T^{ij}_l
&=&-{\cal V}\,\left[2^{n-m}\,i!\,j!\right]^{1/2}\,e^{-\alpha
x_{0}}\,e^{(\alpha_0^{2}/4)}\left[-\alpha_0 \right]^{i-j}\nonumber \\
&&\sum_{k=0}^{j}\frac{\left[ \alpha_0 \right] ^{2k}}{%
2^{k}\,k!\,(k+i-j)!(j-k)!},\,\,i\geq j,  \label{final tl}
\end{eqnarray}
with $T_{l}^{ij}=(-1)^{j-i}\,T_{r}^{ij}$. In the above equation we have defined $\alpha_0=(\sqrt{2}\Delta x_{zp} \,\alpha)$; the matrix elements for the case $i<j$ are calculated by using the hermiticity of the tunneling couplings.
Our model introduces a time-dependent ac-potential, $V(t)=V_{ac}\,\cos(\omega_{ac}t)$, applied between the left and
right quantum dots, and tunneling to and from the leads. Thus, the total Hamiltonian reads: $\hat{H}=\hat{H}_{TQS}+\hat{H}_{ac}+\hat{H}_{leads}+\hat{H}_{dot-leads}$, where $\hat{H}_{ac} =(V_{ac}/2)\,\cos (\omega_{ac}t) \,\left[ \left|
l\right\rangle \left\langle l\right| -\left| r\right\rangle
\left\langle r\right| \right]$ provides the effect of the ac-potential, which is introduced as an
oscillation of opposite phase on the energy levels of the right and left dots flanking the central dot, and
the contribution from the leads is given by
$\hat{H}_{leads}=\sum_{l}\sum_{k_{l}}\epsilon_{k_{l}}\,c_{k_{l}
}^{\dagger }c_{k_{l}}$, where $c^{\dagger}$ and $c$ are creation and destruction operators for an electron with momentum $k_{l}$ in lead $l$, respectively.
Finally, $\hat{H}_{dot-leads}$ stands for the coupling between the external dots and the corresponding leads, and is given by
$\hat{H}_{dot-leads}=\sum_{\{l=L,R\}, \vec{k}} \, V_{l\vec{k}}\,c^{\dag}_{l\vec{k}}\,c_l+h.c.$,
where $V_{l\vec{k}}$ is the dot-lead coupling matrix element, which leads to a coupling strength
$\Gamma_l=2\pi \, \sum_{l,\vec{k}} |V_{l\vec{k}}|^2\, \delta(\epsilon-\epsilon_{l\vec{k}})$.

The equation of motion of the system using the standard master
equation approach for the reduced density matrix (RDM) \cite{Blum},
generalized to include the environment of the oscillator, given
by
$\dot{\rho}_{qs}^{ij}=[\dot{\rho}_{qs}^{ij}]_{RDM}
+[\dot{\rho}_{qs}^{ij}]_{diss}$,
where $\rho_{qs}^{ij}$ represents the matrix element with the dot
states $q$, and $s$ ($q,s=0,l,c,r$), with the oscillator states $i$,
and $j$ ($i,j=0,1,2,...$). The term $[\dot{\rho}_{qs}^{ij}]_{RDM}$
incorporates transitions between the leads and the outer dots,
%
%
\begin{eqnarray}
&&[\dot{\rho}_{qs}^{ij}]_{RDM} =-\frac{i}{\hbar }[ \hat{H},\,\rho ]_{qs}^{ij} \nonumber \\
&&+\left\{
\begin{array}{cc}
\sum_{d\neq q}\left( \Gamma _{d\rightarrow q}\,\rho
_{dd}^{ij}-\Gamma
_{q\rightarrow d}\,\rho_{qq}^{ij}\right) ; \, q=s \\
-\frac{1}{2}\left( \sum_{d\neq q}\Gamma _{q\rightarrow
d}+\sum_{d\neq s}\Gamma _{s\rightarrow d}\right) \,\rho_{qs}^{ij};
\, q\neq s.
\end{array}
\right. \label{masterequation2}
\end{eqnarray}
Here we have also considered the transitions $\Gamma_{r\rightarrow
0}=\Gamma_{0\rightarrow l}=\Gamma$,
%
and the term
$[\dot{\rho}_{qs}^{ij}]_{diss} =-\gamma
[(i+j)/2)]\rho_{qs}^{ij}+\gamma
[(i+1)(j+1)]^{1/2}\rho_{qs}^{i+1,j+1}$,
accounts for the dissipative effects of the oscillator's
environment \cite{Scully} in the low temperature limit; here $\gamma $ stands for the
classical damping rate of the oscillator.
%
The calculations for the current are
performed by numerical integration of the RDM and averaging over the
resulting electronic current measured at the right quantum dot, i.e.
$I=e \,\Gamma \,\sum_i \,[\rho_{rr}^{ii}]_{av}$.


\subsection{Floquet theory for oscillating triple quantum dots}

Since our system is described by a time periodic Hamiltonian, we can use Floquet theory \cite{grifoni,holthaus} in order to analyze
the quasienergy spectrum of the system.
We obtain  the quasienergies by numerical diagonalization of
the Floquet Hamiltonian ${\cal H}=[\hat{H}' -i\hbar\partial /\partial t]$,
(with $\hat{H}'=\hat{H}_{TQS}+\hat{H}_{ac}$),
represented in the composite basis ${\cal D}\bigotimes {\cal O}\bigotimes {\cal T}$ of quantum dots ($|l\rangle ,|c\rangle ,|r\rangle $), oscillator ($i,j=0,1,2,...N$) and the periodic functions of $t$ with period $\tau=(2\pi/\omega_{ac}$), respectively. The matrix elements in this composite representation are given by
\begin{equation}
{\cal H}_{qs}^{ij,n'n}=\langle\langle n' | {\cal H}_{qs}^{ij} |n
\rangle\rangle=\frac{1}{{\tau}}\int_0^{{\tau}}dt
e^{-in'\omega_{ac}t}{\cal H}_{qs}^{ij}e^{in \omega_{ac}t}, \nonumber \\
\end{equation}
where we have used the definition of the inner product in space ${\cal T}$, involving a time average. If $N_t$
is the number of photon sidebands, then $n,n'=0,\pm 1,\pm 2,\pm 3,...,\pm N_t$. This yields,
\begin{eqnarray}
{\cal H}_{ll}^{ij,n'n} &=&(\varepsilon_l+j \hbar
\omega+n\hbar\omega_{ac})\delta_{n'n}\delta_{ij} \nonumber \\
&+&\frac{V_{ac}}{4}(\delta_{n',n+1}+\delta_{n',n-1})
  \delta_{ij};\nonumber \\
{\cal H}_{cc}^{ij,n'n} &=& -\Delta x_{zp}\frac{\varepsilon_b}{2x_0}(
\sqrt{j+1}\,\delta_{i,j+1}+\sqrt{j}\,\delta_{i,j-1}) \delta_{n'n} \nonumber \\
&+&(j\hbar\omega+n\hbar \omega_{ac})
\delta_{ij}\delta_{n'n};\nonumber \\
{\cal H}_{rr}^{ij,n'n} &=&(\varepsilon_r+j \hbar
\omega+n\hbar\omega_{ac})\delta_{n'n}\delta_{ij} \nonumber \\
&-&\frac{V_{ac}}{4}(\delta_{n',n+1}+\delta_{n',n-1})
  \delta_{ij};\nonumber \\
{\cal H}_{lc}^{ij,n'n} &=& {\cal H}_{cl}^{ij,n'n}=T^{ij}_l(\hat{x})\delta_{n'n};\nonumber \\
{\cal H}_{lr}^{ij,n'n} &=& {\cal H}_{rl}^{ij,n'n}=0;\nonumber \\
{\cal H}_{cr}^{ij,n'n} &=& {\cal
H}_{rc}^{ij,n'n}=T^{ij}_r(\hat{x})\delta_{n'n}. \label{elmtxcomplete}
\end{eqnarray}

The Floquet quasienergy spectrum of the system can be obtained by numerical diagonalization of the matrix whose elements are given above.
The eigenvectors of ${\cal H}$ in the composite basis ${\cal D}\bigotimes{\cal O}\bigotimes{\cal T}$ can be written as
\begin{equation}
|\Phi_k(q;j,t)\rangle=\sum_{n,q,j}\,C^k_{nqj}\,e^{-in\omega_{ac}t}|q;j\rangle,
\label{Fourier}
\end{equation}
where
the $kth$ Floquet state, is represented in the composite basis $|q;j\rangle\equiv|q\rangle\bigotimes|j\rangle$,
where the indexes $q$ and $j$ correspond to the dot and oscillator states, respectively.
Thus, the Floquet eigenstates are given in effect by the Fourier coefficients $C^k_{nqj}$.
As we shall later see, the parity of the Floquet states determines the symmetry properties of these Fourier coefficients.

\subsection{Perturbative approach}

In this section we describe an alternative method for exploring the quasienergy spectrum for both static and dynamical systems.
In both cases, we consider that the tunneling $\hat{H}_{tun}$ can be treated as a perturbation, i.e., $T \ll V_{ac}$ \cite{cref_plat,newpert}, allowing the application of standard Rayleigh-Schr\"{o}dinger perturbation theory.
We present here the details of the perturbative treatment for the dynamic case and the static case is given in the Appendix.


We consider for simplicity the case $N=1$ in the oscillator basis ($i,j=0,1$). We start from the Floquet equation corresponding to the
%
Hamiltonian
\begin{equation}
[\hat{H}'-i\hbar\partial /\partial
t]\,|\phi(t)\rangle=\epsilon_i\,|\phi(t)\rangle, \label{eqflo}
\end{equation}
with $\hat{H}'=(\hat{H}_{TQS}+\hat{H}_{ac})$.
The unperturbed Floquet Hamiltonian given by
${\cal H}_{u}=(\hat{H}_{0}+\hat{H}_{ac}+\hat{H}_{osc}-i\hbar\partial /\partial t)$
%
%
is approximately diagonal in the composite basis ${\cal D}\bigotimes {\cal O}$ of the quantum dots and the oscillator (whenever $(\Delta x_{zp}\, \varepsilon_b/2x_0) \ll \hbar \omega$), with matrix elements ${\cal H}_{u,qs}^{ij}\equiv \langle i | {\cal H}_{u,qs}|j\rangle $, given by,
\begin{eqnarray}
{\cal H}_{u,ll}^{ij} &=&\left[(\varepsilon_l+j \hbar
\omega)+\frac{V_{ac}}{2}cos\omega_{ac}t - i\hbar\frac \partial
{\partial t} \right]\delta_{ij}; \nonumber \\
{\cal H}_{u,cc}^{ij} &=& \left[\varepsilon_l-\varepsilon_b/2+j\hbar\omega-i\hbar\frac\partial {\partial t} \right]\delta_{ij}; \nonumber \\
{\cal H}_{u,rr}^{ij} &=&\left[(\varepsilon_r+j \hbar
\omega)-\frac{V_{ac}}{2}cos\omega_{ac}t - i\hbar\frac \partial
{\partial t} \right]\delta_{ij}; \nonumber \\
{\cal H}_{u,lc}^{ij} &=& {\cal H}_{u,cl}^{ij}=0;\nonumber \\
{\cal H}_{u,lr}^{ij} &=& {\cal H}_{u,rl}^{ij}=0;\nonumber \\
{\cal H}_{u,cr}^{ij} &=& {\cal H}_{u,rc}^{ij}=0,
\label{elemmatriz}
\end{eqnarray}
with $i,j=0,1$. For the particular case of a symmetrical TDQS the unperturbed Hamiltonian ${\cal H}_u$ is exactly diagonal in the composite basis ${\cal D}\bigotimes {\cal O}$, and no approximations are needed.

Following the same procedure as in the static triple dot case discussed in the Appendix, we obtain the following
eigenvectors,
\begin{eqnarray}
|\phi_{0l}^u \rangle &=&
e^{-i(\varepsilon_b/2-\epsilon_{0l})t/\hbar}\,e^{-i\xi
\sin \omega_{ac}t}|0\,l\rangle; \nonumber \\
|\phi_{1l}^u \rangle &=&
e^{-i(\varepsilon_b/2-\epsilon_{1l})t/\hbar}\,e^{-i\xi
\sin \omega_{ac}t}|1\,l\rangle; \nonumber \\
|\phi_{0c}^u \rangle &=& e^{-i\epsilon_{0c}t/\hbar}|0\,c\rangle; \nonumber \\
|\phi_{1c}^u \rangle &=& e^{-i\epsilon_{1c}t/\hbar}|1\,c\rangle; \nonumber \\
|\phi_{0r}^u \rangle &=&
e^{-i(\varepsilon_b/2-\epsilon_{0r})t/\hbar}\,e^{-i\xi \sin
\omega_{ac}t}|0\,r\rangle; \nonumber \\
|\phi_{1r}^u \rangle &=&
e^{-i(\varepsilon_b/2-\epsilon_{1r})t/\hbar}\,e^{-i\xi \sin
\omega_{ac}t}|1\,r\rangle,  \label{eigen3}
\end{eqnarray}
where $\xi=V_{ac}/2\omega_{ac}$.

The corresponding Floquet eigenenergies are given by,
\begin{eqnarray}
\epsilon_{0l} &=&\varepsilon _l+n^{(0)}\hbar \omega _{ac};  \nonumber
\\
\epsilon_{1l} &=&\varepsilon _l+\hbar \omega +n^{(1)}\hbar \omega
_{ac};  \nonumber \\
\epsilon_{0c} &=&\varepsilon _l-(\varepsilon _b/2)+n^{(3)}\hbar \omega
_{ac};  \nonumber \\
\epsilon_{1c} &=&\varepsilon _l-(\varepsilon _b/2)+\hbar \omega
+n^{(4)}\hbar \omega _{ac};  \nonumber \\
\epsilon_{0r} &=&\varepsilon _r+n^{(5)}\hbar \omega _{ac};  \nonumber
\\
\epsilon_{1r} &=&\varepsilon _r+\hbar \omega +n^{(6)}\hbar \omega
_{ac},  \label{floq}
\end{eqnarray}
where $n^{(k)}=0,\pm 1,\pm 2,....$, with $k=1,...,6$.

We now proceed to calculate the matrix elements of the perturbation
$\hat{H}_{tun}$ in the basis $|\phi^u_{i,q} \rangle$, that is, we compute
$P_{qs}^{ij}=\langle\langle \phi^u_{i,q}|\hat{H}_{tun}|\phi^u_{j,s}
\rangle\rangle$, where $i,j=0,1$, and $q,s=l,c,r$.
We can verify that
$P_{ll}^{ij}=P_{cc}^{ij}=P_{rr}^{ij}=P_{lr}^{ij}=P_{rl}^{ij}=0$, while the other non-zero matrix
elements can be calculated by using the identity given by equation (\ref{identbess}), and by
considering two important conditions: (i) that  an alignment of the
ac-sidebands occurs i.e.
$\epsilon_{jc}=\epsilon_{jl}$, and $\epsilon_{jc}=\epsilon_{jr}$, whenever $\varepsilon_b=2s' \hbar \omega_{ac}$, (as in the static case), and that (ii)  $\hbar \omega=n'\hbar \omega_{ac}$ (where $s'$ and $n'$ are positive integers);
the $6\times 6$ matrix $P\equiv(P_{qs}^{ij})$ is then given by

%
%
\begin{equation}
P=\left[
\begin{array}{ccc}
\left( P_{ll}\right)  & \left( P_{lc}\right)  & \left( P_{lr}\right)  \\
\left( P_{cl}\right)  & \left( P_{cc}\right)  & \left( P_{cr}\right)
\\
\left( P_{rl}\right)  & \left(P_{rc}\right)  & \left( P_{rr}\right)
\end{array}
\right],
\label{lap}
\end{equation}
where the $2\times 2$ submatrices are
\begin{equation}
\left(P_{lc}\right)=\left[
\begin{array}{ll}
T_l^{00}J_{-s'}(\xi) & T_l^{01}J_{n'-s'}(\xi ) \\
T_l^{10}J_{-n'-s'}(\xi ) & T_l^{11}J_{-s'}(\xi )
\end{array}
\right];  \label{pcl}
\end{equation}
\begin{equation}
\left(P_{cl}\right)=\left[
\begin{array}{ll}
T_l^{00}J_{-s'}(\xi ) & T_l^{01}J_{-n'-s'}(\xi ) \\
T_l^{10}J_{n'-s'}(\xi ) & T_l^{11}J_{-s'}(\xi )
\end{array}
\right];  \label{pcl}
\end{equation}
\begin{equation}
\left(P_{cr}\right)=\left[
\begin{array}{ll}
T_r^{00}J_{s'}(\xi ) & T_r^{01}J_{s'-n'}(\xi ) \\
T_r^{10}J_{n'+s'}(\xi ) & T_r^{11}J_{s'}(\xi )
\end{array}
\right];
\end{equation}
\begin{equation}
\left(P_{rc}\right)=\left[
\begin{array}{ll}
T_r^{00}J_{s'}(\xi ) & T_r^{01}J_{n'+s'}(\xi ) \\
T_r^{10}J_{s'-n'}(\xi ) & T_r^{11}J_{s'}(\xi )
\end{array}
\right];
\end{equation}
\begin{equation}
\left(P_{ll}\right)=\left(P_{cc}\right)=\left(P_{rr}\right)=\left(P_{lr}\right)=\left(P_{rl}\right)=\left[
\begin{array}{ll}
0 & 0 \\
0 & 0
\end{array}
\right].
\end{equation}
The quasienergies are computed by a numerical diagonalization of $P$ (equation (\ref{lap})); in the case of the symmetrical ($\varepsilon_b=0$) TDQS, with $\omega_{ac}=0.2$, and $\omega=1$, we get $s'=0$ and $n'=5$.
For the asymmetrical case ($\varepsilon_b\neq 0$), and the particular case of $\varepsilon_b=0.4$, we have $s'=1$
in equation (\ref{lap}).

For the static case ($\hat{H}_{osc}=0$) we obtain the following analytical expressions for the quasienergies (the details are discussed in the Appendix),
\begin{eqnarray}
\epsilon_l &=& \left[\frac{\varepsilon_b}{2}-\sqrt{2}\,T\,J_{s'}(\xi) \right] \,(\mbox{mod} \,\, \hbar \omega_{ac}); \nonumber \\
\epsilon_r &=& (-\epsilon_l)\, (\mbox{mod} \,\, \hbar \omega_{ac}); \nonumber \\
\epsilon_c &=&0 \, (\mbox{mod} \,\, \hbar \omega_{ac}); \label{perturb_stat}
\end{eqnarray}
where $T=-{\cal V}\exp(-\alpha x_0)$.
%
This result is in agreement with a general perturbative treatment of
ac-driven tight-binding chains of quantum wells or quantum dots
\cite{crefPRB2004,longhi,villas-boas}.
Furthermore, it can be seen that an exact crossing between all three quasienergies is
obtained whenever $\varepsilon_b$ is an {\it even} multiple of
$\hbar\omega_{ac}$ and it corresponds to  the roots of the
corresponding Bessel function $J_{s'}$ (See for example figure \ref{statcuasiener} (b), where, for the chosen parameters, $s'=1$).


\section{\label{results}Results}

In what follows, we shall explore the main features of the current
at finite detuning, and the quasienergy spectrum for the TDQS driven by an ac gate
voltage in the weak interdot coupling regime, with special emphasis on the CDT phenomenon.
All of the calculations are worked out in units where
$\hbar=2m=e=1$.
\begin{figure}[!h]
\center
\rotatebox{0}{\includegraphics[width=3.3in]{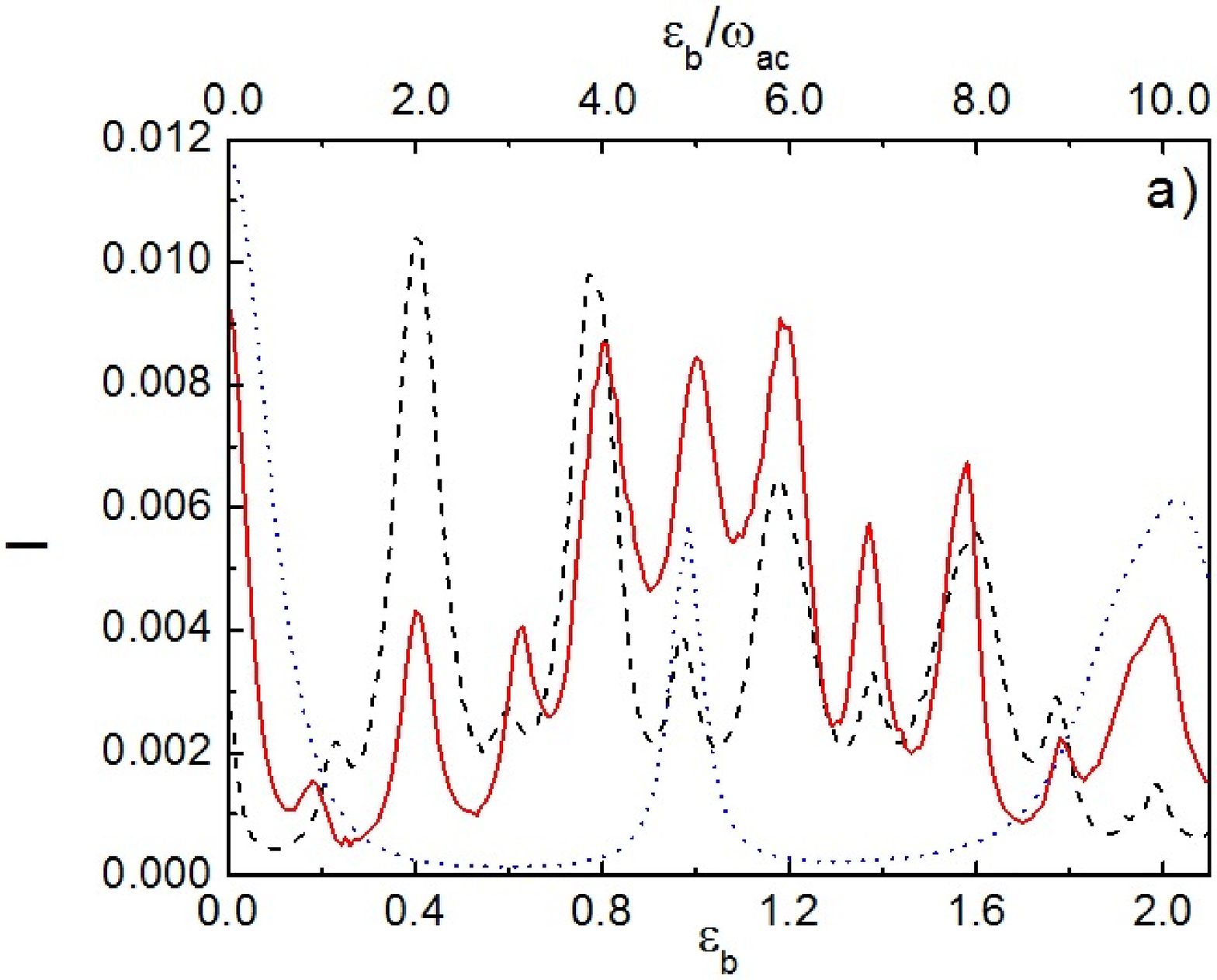}}
\rotatebox{0}{\includegraphics[width=3.3in]{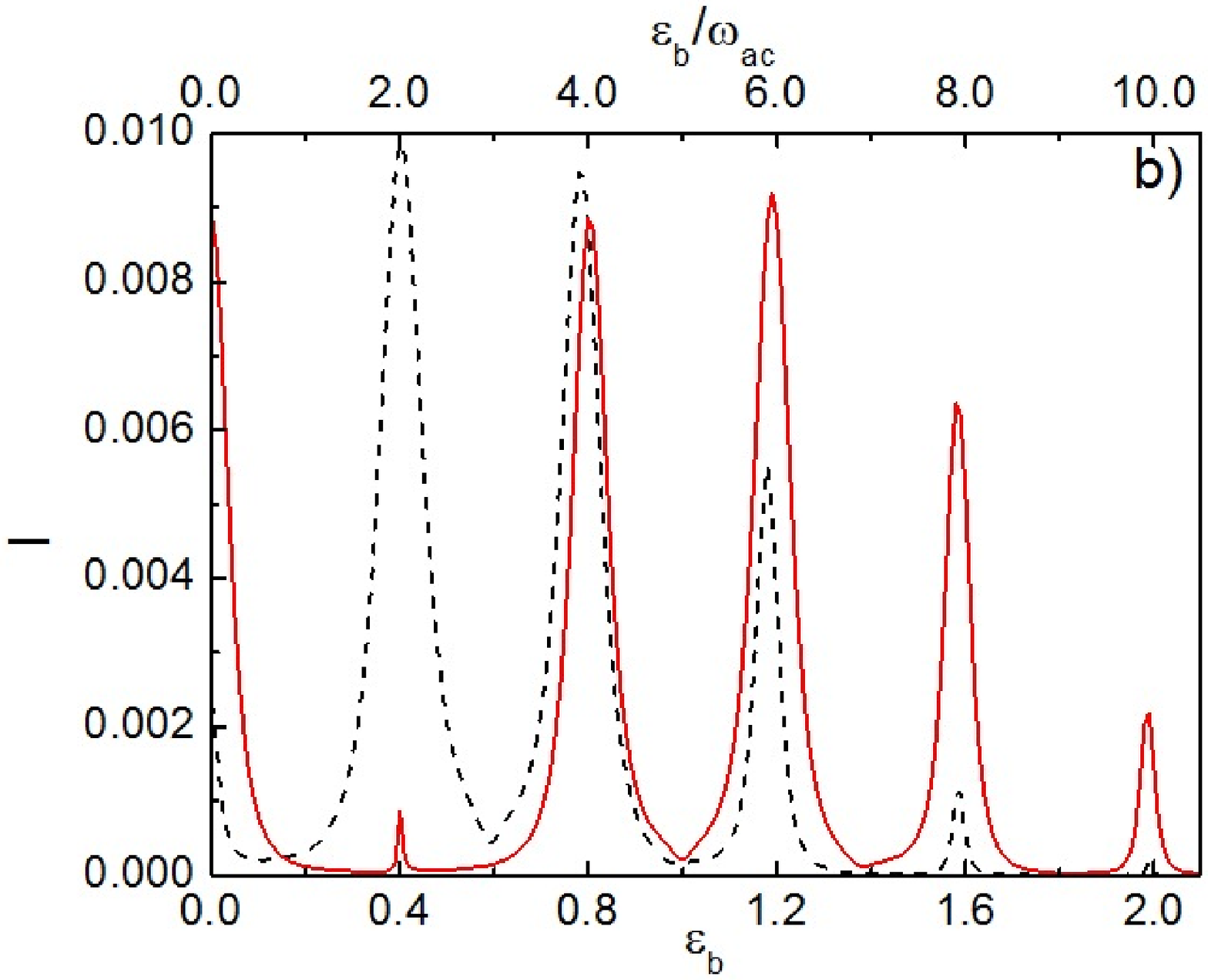}}
\caption{$I$ versus $\varepsilon_b$ for a triple dot system for the ac parameters $\omega_{ac}=0.2$, $V_{ac}=1$ (black dashed line) , $V_{ac}=1.533$ (red solid line), corresponding to first root of $J_1(\xi$), and typical values of the parameters \{$V$; $\gamma$; $x_0$; $\Gamma$; $\alpha$\}=\{$0.5$; 0.01; $5.0$;
$0.05$; $0.4$\}, for both the (a) dynamic ($\omega=1=5 \omega_{ac}$), and (b) static ($\omega=0$) cases.
In (a) we also include for comparison the undriven current ($V_{ac}=0$, blue dotted line).
Note the reduction of the resonance peak at $\varepsilon_b=0.4$ (red solid line) in both cases, and the appearance of resonances at odd multiples of $\omega_{ac}$ in (a). These are due to the overlap between wave functions of left and right quantum dots as discussed in Ref. \cite{APSpaper}.} \label{current}
\end{figure}
\begin{figure}[!tbp]
\center
\rotatebox{0}{\includegraphics[width=3.3in]{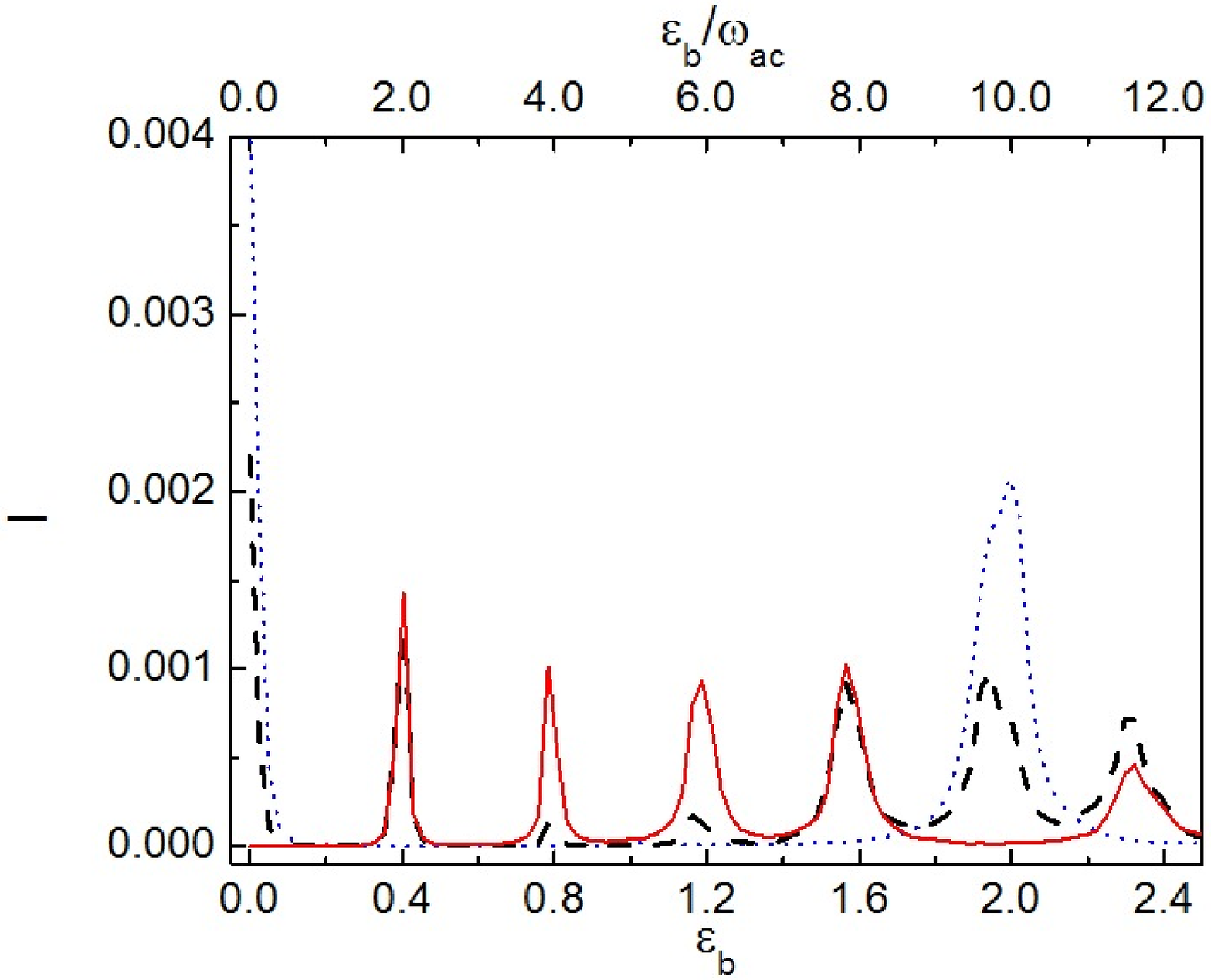}}
\caption{$I$ versus $\varepsilon_b$ for a triple dot system in the dynamic case for a smaller tunneling than in the previous case of figure \ref{current} (a), by choosing  $\alpha=0.8$ with ac parameters $\omega_{ac}=0.2$, $V_{ac}=0.5$ (black dashed line) and $V_{ac}=0.96192$ (red solid line), corresponding to first root of $J_0(\xi)$ ($\xi=2.4048$), and values of the parameters \{$V$; $\gamma$; $x_0$; $\Gamma$\}=\{$0.5$; 0.01; $5.0$; $0.05$\}. Also shown is the current for the undriven case (blue dotted line). We can see that the undriven current exhibits two well defined resonances at $\varepsilon_b=0$, and $2.0$. These resonances are  effectively turned off by adequately tuning the field intensity $V_{ac}$ to a CDT condition.} \label{current2}
\end{figure}
In figure \ref{current} (a) we present the electronic current, $I$, as a function of detuning $\varepsilon_b$, where a series of resonances at even and odd multiples of the ac-frequency $\omega_{ac}$ can be seen.
In a previous work \cite{APSpaper}, we analyzed numerically the tunneling current through a Triple Quantum Dot (TQD) oscillating device and found that the positions of these peaks obeyed sum rules indicating the number of photons emitted or absorbed in the process ($\nu$) and the final oscillator state ($n$) (($\nu$,$n$) for abbreviation).
In figure \ref{current} (a) we present results for the current as a function of detuning $\epsilon_b$, for two values of the intensity of the ac-field $V_{ac}$. For $V_{ac}=1$ (black dashed line), we note
that  there are current resonances at values of $\varepsilon_b$ which are integer multiples of $\omega_{ac}$.
Stronger peaks are observed at even multiples of $\omega_{ac}$
corresponding to the alignment of ac-sidebands developing from eigenstates $\epsilon_l,\epsilon_c$
and $\epsilon_r$. The weaker peaks at odd multiples of $\omega_{ac}$, correspond to the alignment of levels $\epsilon_l$
and $\epsilon_r$.
These are due to the overlap of left and right quantum dots, as discussed in Ref. \cite{APSpaper}.

Let us recall under what conditions the electronic current exhibits the CDT phenomenon.
The current peak at $\varepsilon_b=0.4$ (and $\omega_{ac}=0.2$), corresponds to $\nu=1$
emitted photons and $n=0$ is the final oscillator state (see Ref. \cite{APSpaper}).
Interestingly, if we calculate the current for $\xi=3.832$
($V_{ac}=1.533$, red solid line in figure \ref{current} (a))
corresponding to the first root of $J_1(\xi)$, with all other
parameters fixed, we see that the peak at $\varepsilon_b=0.4$
decreases significantly. This is the phenomenon of
coherent destruction of tunneling in our triple dot
quantum shuttle device. Additionally, we note that at
$\varepsilon_b=2.0$, the opposite situation occurs, where the
resonance for $V_{ac}=1$ (dashed black line) is small compared to the case
$V_{ac}=1.533$ (red solid line). The reason is that this peak
corresponds to $\nu=0, n=1$ and $V_{ac}=1$ is very near the value
corresponding to the first zero of $J_0(\xi)$ ($V_{ac}=0.96192$, corresponding to $\xi=2.4048$).
Included in this figure for comparison purposes is the current for the undriven case (blue dotted line), where resonances appear at multiples of the oscillator frequency $\omega=1$.
In figure \ref{current} (b), we present the results for the static case ($\omega=0$). Here we see that current resonances appear only at values of
$\varepsilon_b$ which are {\it even} multiples of $\omega_{ac}$.
These resonances correspond to $\nu=1,2,3,...$ emitted photons involving the final oscillator state $n=0$ as expected.
Note that the peak at $\varepsilon_b=0.4$ practically disappears at the CDT condition (red solid line).
Thus, for this particular choice of parameters, the CDT phenomenon is more evident in the static case than in the dynamic case,
where additional tunneling channels (besides $\nu=1$) contribute to the current.
This will be corroborated further when we analyze below the properties of the quasienergy spectrum. Interestingly, we also note the current peaks at $\varepsilon_b=1.6$ and $\varepsilon_b=2.0$ for $V_{ac}=1$ (black dashed line) have also decreased. According to the sum rules, these peaks correspond to $\nu=4$ and $\nu=5$ emitted photons, respectively, i.e. interdot tunneling is renormalized by Bessel functions $J_4(\xi)$ and $J_5(\xi)$ which, for  $V_{ac}=1, \xi=2.5$, are small ($\lesssim 10^{-2}$).

Next, in figure \ref{current2} we plot the electronic current $I$ as a function of detuning $\varepsilon_b$ in a TQD oscillating system (i.e. dynamic case with $\omega=1$) for smaller tunneling couplings $T_l$ and $T_r$ (i.e. smaller than in the case of figure 2 (a)), by choosing a value of $\alpha=0.8$. The undriven current (blue dotted line) exhibits two strong shuttling resonances at values of $\varepsilon_b=0$, and $2.0$, which correspond to alignment of the energy levels of the vibrational subbands of the three dots. When the ac voltage is applied (black dashed line), an additional energy level splitting occurs which gives rise to the new resonances at even multiples of $\omega_{ac}$. By an adequate manipulation of the driving field intensity, we can induce the CDT condition (red solid line) in order to turn off the shuttling resonance peaks corresponding to $\varepsilon_b=0$, and $2.0$. Also note there are no resonances at odd multiples of the driving frequency. The reduction in tunneling effectively cancels the overlap between functions of left and right quantum dots.

\begin{figure}[!tbp]
\center \rotatebox{0}{\includegraphics[width=3.3in]{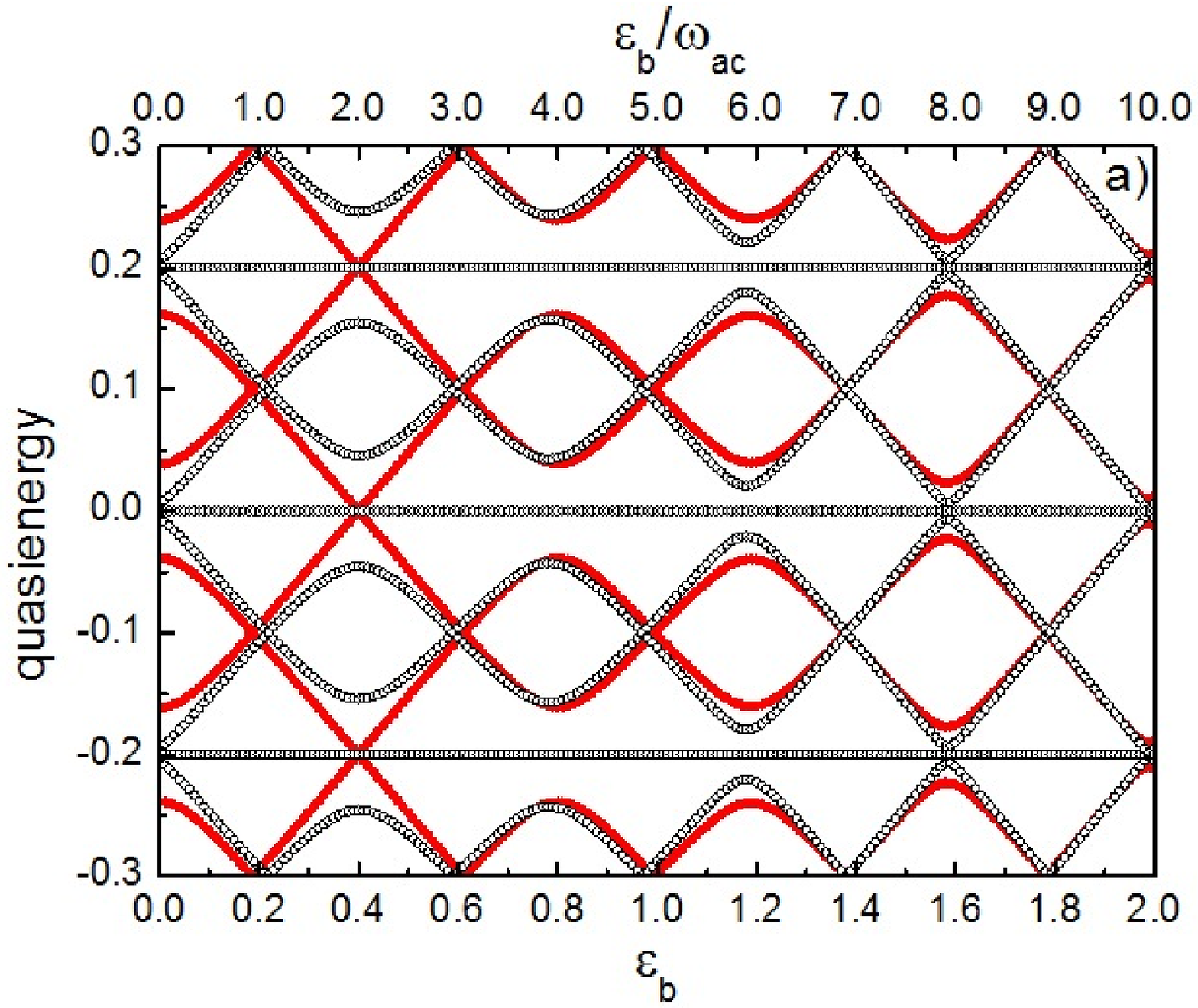}}
\rotatebox{0}{\includegraphics[width=3.3in]{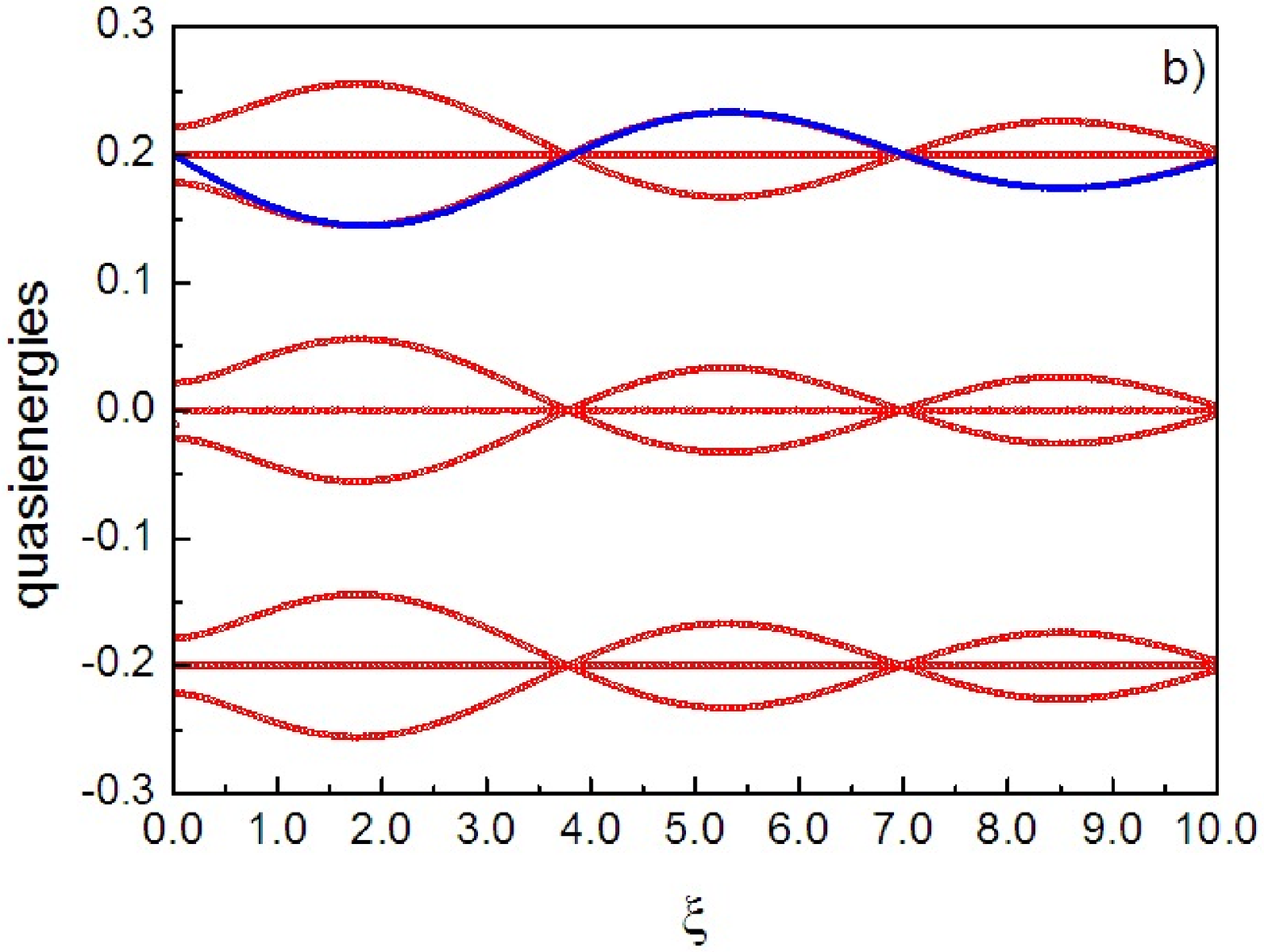}}
\caption{Static case: (a) Quasienergy versus $\varepsilon_b$ for ac
parameters $\omega_{ac}=0.2$, and $V_{ac}=1$ (black hollow dot line)
and $V_{ac}=1.533$ (red solid dot line), and typical values of the
parameters \{$V$; $x_0$; $\Gamma$; $\alpha$\}=\{$0.5$; $5.0$;
$0.05$; $0.4$\}. Note the crossings of the red line at
$\varepsilon_b=0.4$, and of the black line at $\varepsilon_b=1.6$, and $2.0$,
both indicative of CDT. (b) Quasienergy versus
$\xi=(V_{ac}/2\omega_{ac})$ for $\varepsilon_b=0.4$, and same
parameters as in (a) (red hollow dot line). Results from a first
order perturbation calculation given by equation
(\ref{perturb_stat}) (blue solid line), are also shown.} \label{statcuasiener}
\end{figure}
\begin{figure}[!tbp]
\center
\rotatebox{0}{\includegraphics[width=3.3in]{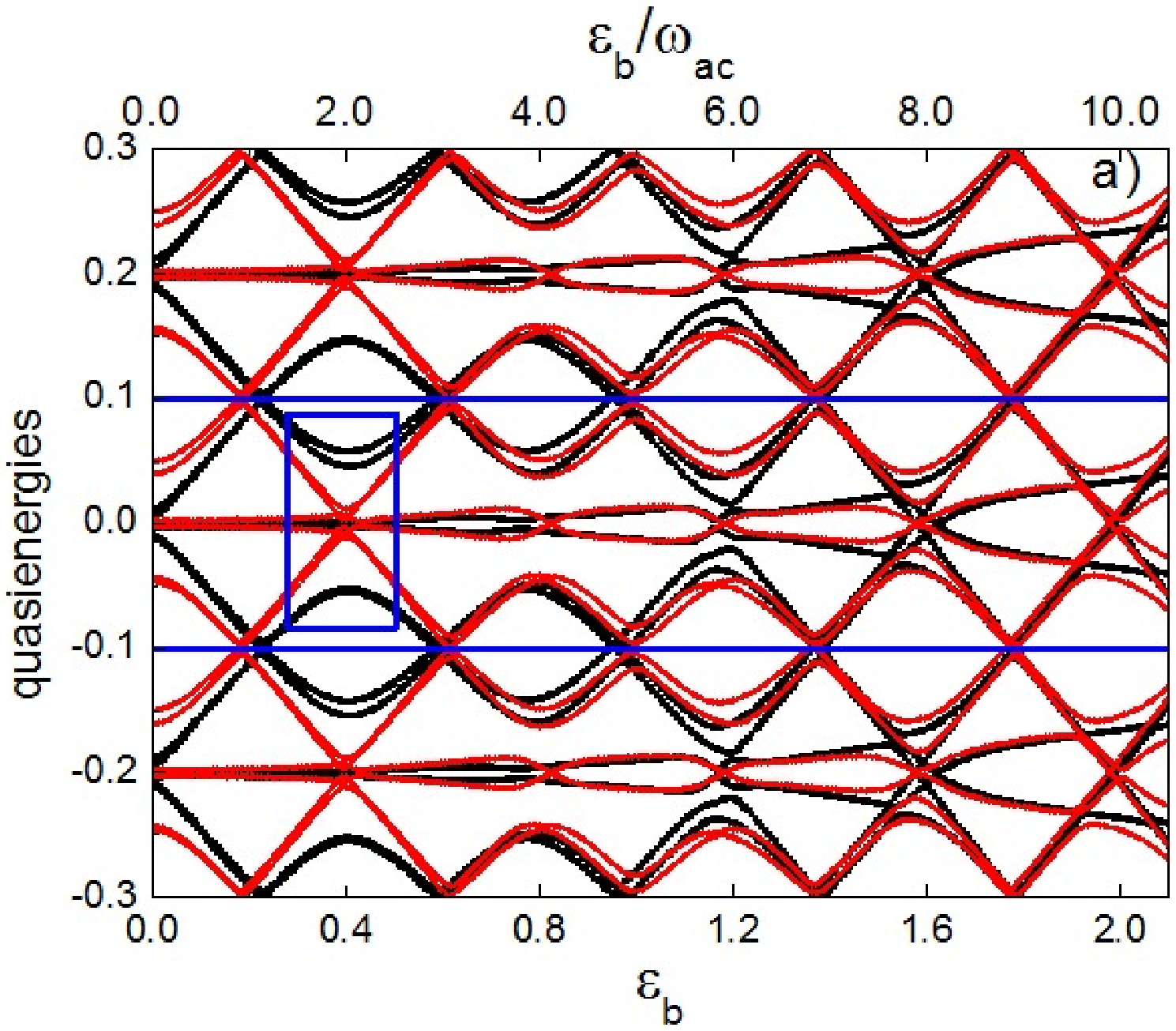}}
\rotatebox{0}{\includegraphics[width=3.3in]{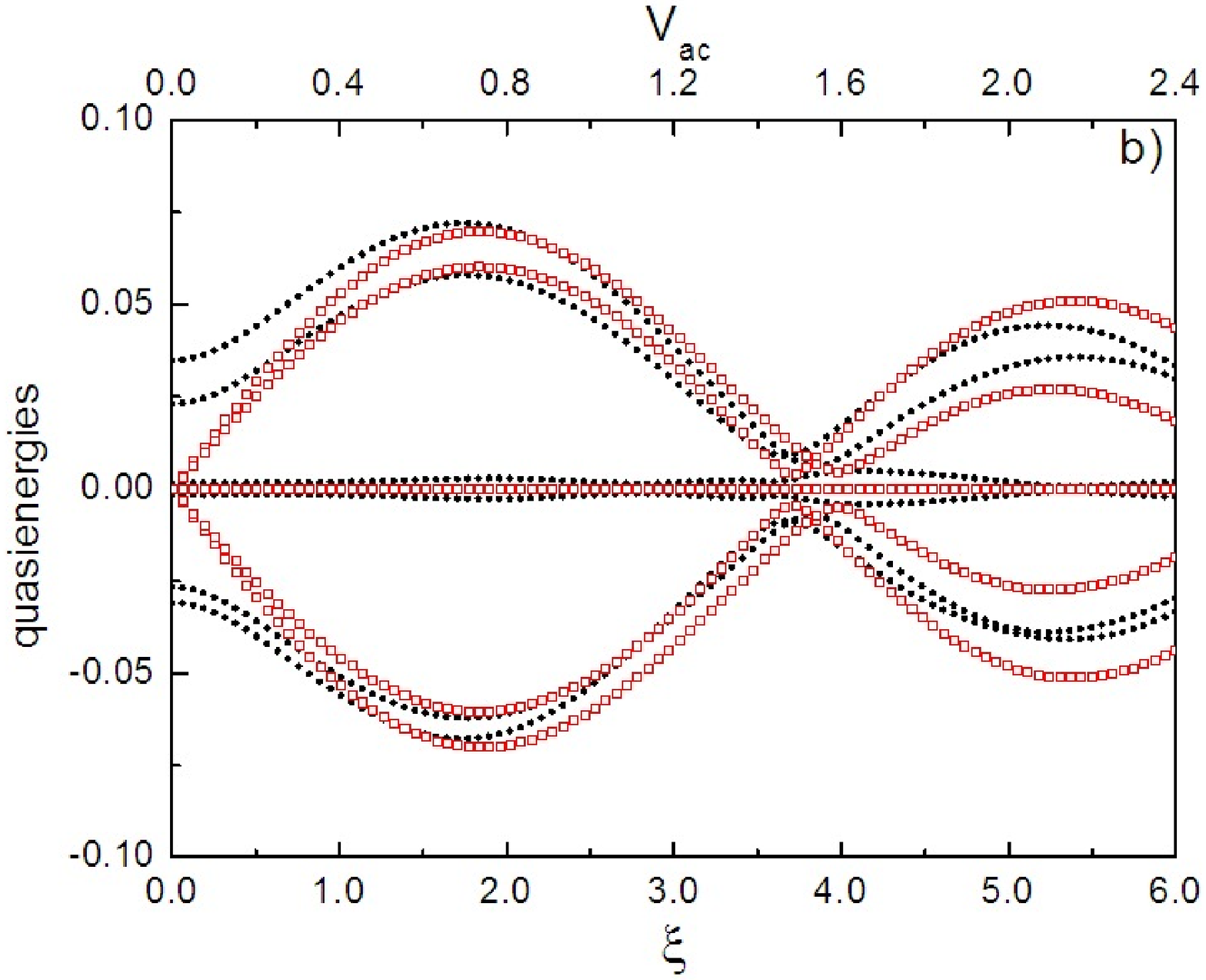}}
\caption{Dynamic case: (a) Quasienergies as a function of detuning $\varepsilon_b$ for $V_{ac}=1$ (black solid dot line) and $V_{ac}=1.533$ (red solid dot line), corresponding to  CDT conditions for $\varepsilon_b=0.4$, with $\omega_{ac}=0.2$ and typical values of the parameters $\{V$; $x_0$; $\Gamma$; $\alpha\}$=$\{0.5$; $5.0$; $0.05$;
$0.4\}$ with $N_t=20$. (b) Quasienergies as a function of $\xi=(V_{ac}/2\omega_{ac})$ for
$\varepsilon_b=0.4$ with $\omega_{ac}=0.2$ and the rest of the parameters the same as in (a) (black solid dot line). The result of a first order perturbative calculation obtained from equation (\ref{lap}) (red hollow square line) is included for comparison.}
\label{dyncuasiener}
\end{figure}
%

%
\begin{figure}[!tbp]
\center
\rotatebox{0}{\includegraphics[width=3.3in]{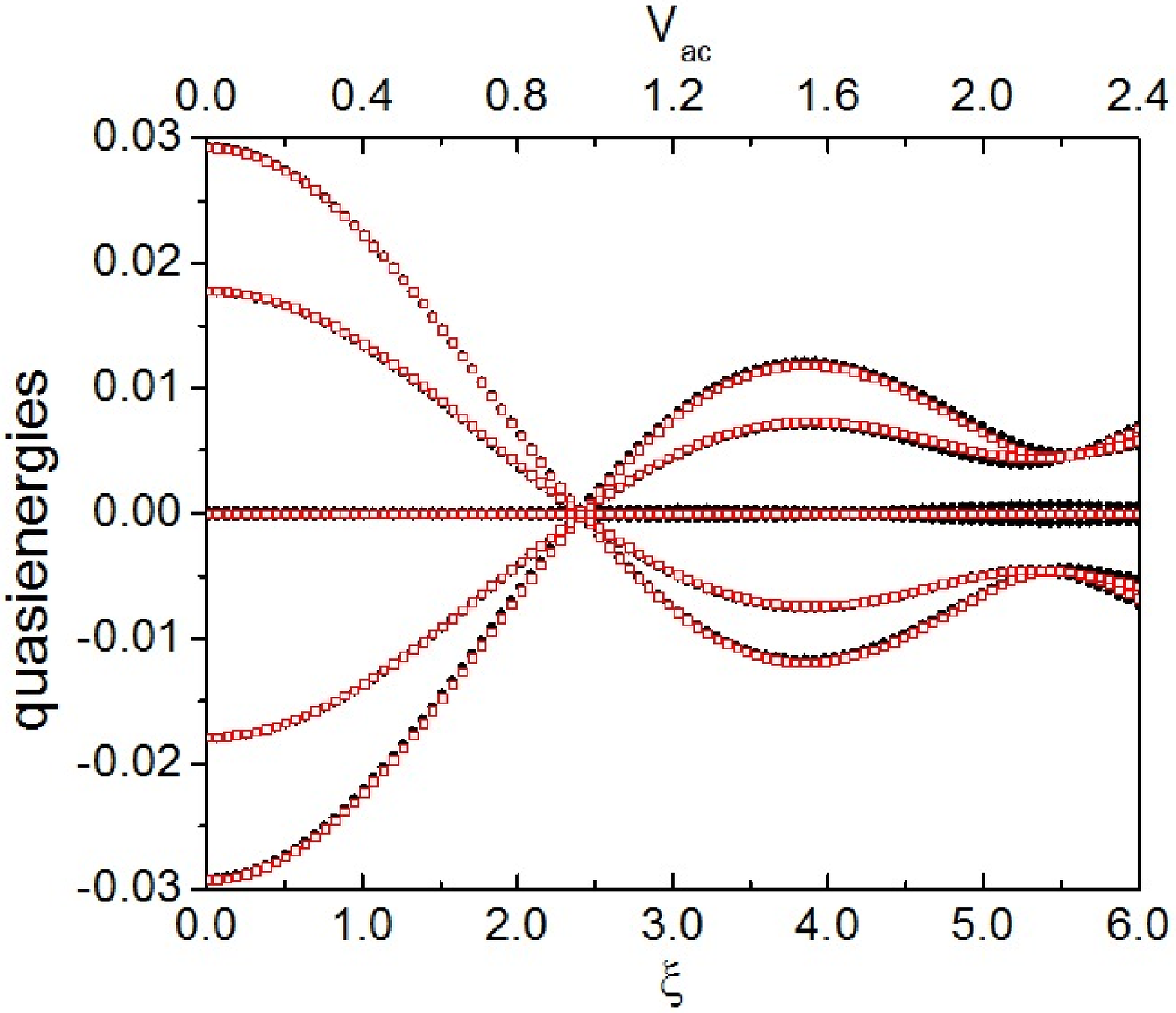}}
\caption{Quasienergies as a function of $\xi=(V_{ac}/2\omega_{ac})$ for
$\varepsilon_b=0$ with $\omega_{ac}=0.2$ and typical values of the parameters $\{V$; $x_0$; $\Gamma$\}$=$\{0.5$; $5.0$; $0.05$ \}$, with $N_t=20$ (black solid dot line), for the case $\alpha=0.8$ of figure \ref{current2}. Note the excellent agreement obtained as compared with the result of a first order perturbative calculation obtained from equation (\ref{lap}) (red hollow square line), where both plots almost completely overlap.}
\label{dyncuasiener_new}
\end{figure}
%

\subsection{Quasienergy spectrum analysis}

Next, we carry out a numerical calculation of the quasienergies
as a function of detuning $\varepsilon_b$ and driving field parameter $\xi$. Following Fromherz \cite {fromherz} and Je \etal \cite{coreanos}, we use $N_t=20$ Fourier components in
order to get accurate results for the range of values of the driving
field intensity $V_{ac}$.

In the static case, when we examine the quasienergies as a function of detuning $\varepsilon_b$ (figure \ref{statcuasiener} (a)) we see the three
quasienergies $\epsilon_l, \epsilon_c, \epsilon_r$, along with their
replicas at multiples of $\omega_{ac}$ ($V_{ac}=1.0$, black hollow dot line), in agreement with the
Floquet theorem. Furthermore, we see anticrossings involving all
three levels at the values of $\varepsilon_b$ where current
resonances appear.
For $V_{ac}=1.533$ (red solid dot line) corresponding to the first zero of $J_1(\xi)$ (CDT condition), an exact crossing develops at $\varepsilon_b=0.4$, in agreement with the disappearance of the current resonance at this detuning (figure \ref{current} (b)).
Two more CDT conditions occur at $\varepsilon_b=1.6$ and $\varepsilon_b=2.0$, indicated by the crossing of the black hollow dot line at these values, as discussed before.
Also for the static case, in figure \ref{statcuasiener} (b) we show the quasienergies as a function of $\xi$ for the case $\varepsilon_b=0.4$.
Note the exact crossings (CDT) at $\xi \approx 3.832$ and $7$,
corresponding to the first two roots of the Bessel function
$J_1(\xi)$, and to the quenching of the current resonance at this detuning in figure 2 (b). We see that
the numerical results agree quite well with the perturbation
calculation (equation (\ref{perturb_stat})), using $J_1(\xi)$.

For the dynamic case, in figure \ref{dyncuasiener} (a) we show results for the quasienergies as a function of detuning $\varepsilon_b$, for the same two values of the ac-field intensity. For $V_{ac}=1$ (black dot line) we see a pattern of crossings (anticrossings) where $\varepsilon_b$ is an odd (even) multiple of $\omega_{ac}$ where current resonances appear. Interestingly, for $V_{ac}=1.533$ (red line in figure \ref{dyncuasiener} (a)), corresponding to the CDT condition, we note a
drastic reduction in the anticrossing separation (compared with the previous case $V_{ac}=1$) at $\varepsilon_b=0.4$ (the region highlighted by the blue square in figure \ref{dyncuasiener} (a)), which signals the appearance of CDT. Thus, even for the case of  finite detuning, where no symmetry properties can be invoked for the eigenstates, the presence of CDT can still be identified with these near crossings in the quasienergy spectrum. This is further corroborated by plotting the quasienergies as a function of $\xi$ (figure \ref{dyncuasiener} (b)), for fixed $\varepsilon_b=0.4$, where again we obtain these near crossings at the roots of the Bessel function $J_1(\xi)$. Also plotted in this figure are the results of the perturbation calculation, equation (\ref{lap}), using $J_1(\xi)$ (red hollow squares). Note that the perturbation calculation also reproduces the near crossing.
%
In figure \ref{dyncuasiener_new} we show the quasienergies as a function of $\xi$ (black solid dot line), for the case of a smaller value of tunneling ($\alpha=0.8$) compared with the case discussed in figure \ref{current2}, for $\varepsilon_b=0$ (symmetrical system), and $\omega_{ac}=0.2$. In this case the CDT phenomenon occurs at a particular value of $V_{ac}=0.96192$, with a ratio $\xi=2.4048$ that corresponds to the first zero of $J_0(\xi)$. The perturbation calculation using equation (\ref{lap}) (red hollow square line) gives an excellent agreement with the numerical results.

For completeness, let us explore more features of the CDT phenomenon in TQD oscillating symmetrical systems ($\varepsilon_b=0$), by considering the case presented in figure \ref{dyncuasiener} (a).
\begin{figure}[!tbp]
\center
\rotatebox{0}{\includegraphics[width=3.5in]{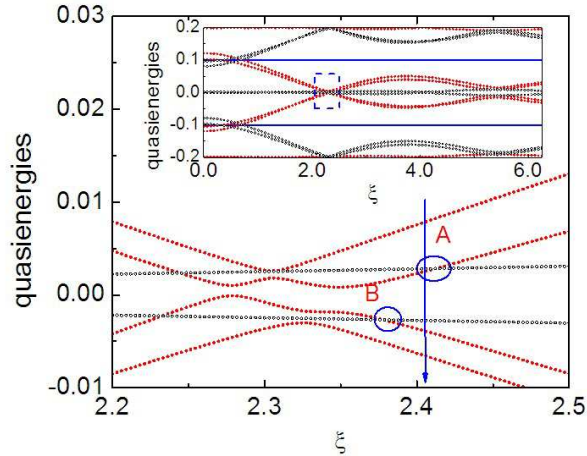}}
\caption{The inset shows the quasienergies in the first Brillouin zone (delimited with blue solid lines to help the eye) as a function of $\xi$, for
$\varepsilon_b=0$, with fixed $\omega_{ac}=0.2$, $N=1$ oscillator states, $N_t=20$ and the rest of the parameters
the same as in figure \ref{dyncuasiener} (a).
The main graph is an amplification of the quasienergies in the vicinity of $\xi=2.4048$ (indicated by a blue arrow) which corresponds to the zero of $J_0(\xi)$ where CDT occurs. The regions $A$, and $B$ (see text) exhibit two crossings of the quasienergies. The quasienergies are classified according to parity i.e. even states (red solid dot line) and odd states (black hollow dot line).}
\label{cuasiener}
\end{figure}
%
%
We perform a Floquet analysis of the CDT by analyzing the quasienergy spectrum of the
system.
For our calculations, we choose $N=1$, i.e. two oscillator states $n=0,1$ for each quantum dot.
In figure \ref{cuasiener} we plot the quasienergies as a function of the ratio $\xi$.
The inset of figure \ref{cuasiener} shows the quasienergies in the first Brillouin zone ($-\omega_{ac}/2 \le \epsilon_{\alpha} \le \omega_{ac}/2$) as a function of
$\xi$, where we can appreciate a complex pattern of crossings and anticrossings of the quasienergies in the vicinity of $\xi=2.4048$; an amplification of the region where this behavior occurs is illustrated in the main graph. Here we can identify regions $A$ and $B$, where two crossings of the quasienergies take place.  The corresponding eigenstates exhibit opposite parities, where the even states are denoted by red solid dots and the odd states with hollow black dots.
The above results are supported by detailed numerical studies, and by using an interesting symmetry property of the Floquet Hamiltonian. At zero detuning,
the Floquet Hamiltonian is invariant under a generalized parity transformation $S_{GP}: (x,t)\rightarrow (-x,t+\tau /2)$ \cite{hanggi98,kohler}. Hence, the Floquet states can be classified according to \emph{parity}. From equation (\ref{Fourier}), we can write the Fourier coefficients, $C_{n,q,j}^k$, as
\begin{equation}
C_{n,q,j}^k=\frac{1}{\tau}\int^{\tau}_0 \langle q;j | \Phi_k \rangle e^{in \omega_{ac}t} dt,
\label{coefficients}
\end{equation}
and, after performing the $S_{GP}$ transformation, it can be shown that the above coefficients obey the following relations,
\begin{equation}
C_{n,q,j}^k=\pm(-1)^n(-1)^j\,C^k_{n,-q,j},
\label{evenodd}
\end{equation}
where the upper (lower) sign stands for even (odd) states, $k$ stands for the $k$-th Floquet state, and the indexes are defined as before: $n$, $q$, and $j$, associated to the temporal, dot, and oscillator basis, respectively.
In our notation, the $-q$ index in the right hand side of the above equation indicates a spatial site inversion of the dots, i.e. $l \leftrightarrow r$.
Using these properties, we have verified that at $A$ and $B$ (figure \ref{cuasiener}) the corresponding eigenstates exhibit opposite parities.


\section{\label{conclusions}Conclusions}

In summary, we have carried out a Floquet analysis in an {\it  oscillating} ac-driven triple quantum dot system in order to investigate the quasienergy spectrum as a function of the field parameters and the external voltage.  The CDT condition is discussed  in terms of the properties of the quasienergy spectrum as a function of detuning and ac parameters. For that purpose, in the strong driving regime,  we extend a perturbative analysis for the static case to the dynamic configuration.
For finite detuning, we find conditions under which, CDT is realized in both static and dynamic cases.
In the latter case, a smaller reduction of the current resonance peak is obtained as compared with the static case, due to additional channels contributing to the tunneling current.
We have also found in the dynamic case that the ac drive, for certain field parameters, induces a CDT condition that destroys the tunneling current peaks (tunneling channels) for the undriven oscillator. Therefore, ac fields are able to modify and even block the tunneling current otherwise flowing through an oscillating triple quantum dot.
From the analysis of current-voltage characteristics,
 we obtain current resonances at
voltages which are integer multiples of the driving frequency
$\omega_{ac}$. For the case of even
multiples of $\omega_{ac}$, in the static case, the numerical results for the quasienergies can be reproduced
by analytical perturbative calculations. 
Exact crossings of the quasienergy spectrum are obtained, characterizing the CDT condition, while near crossings are obtained in the dynamic case.
For the case of odd
multiples of $\omega_{ac}$, the numerical result cannot be
reproduced by first order perturbative calculations.
With respect to this case, it is important to stress that interdot tunneling is coherent in the present model and, therefore,  tunneling is included to all orders in our numerical calculation. However, this is not the case for the perturbative model presented here, where we have only considered  first order perturbation theory.
For zero detuning, by exploiting the invariance of the Hamiltonian of the system under a generalized parity transformation, we have found that the Fourier components of the Floquet states satisfy analytical expressions which take into account the parity of the oscillator states. These allow us to classify the quasienergy spectrum
of the oscillating driven TQD according to parity. By using these expressions, we show that, at zero detuning,  in the vicinity of the CDT condition, the quasienergies exhibit crossings, where the eigenvectors are characterized by opposite parities.
Our results are a contribution in ongoing efforts to control electron transport in few electron nanoelectromechanical quantum dot arrangements. They should be relevant, for example, in Molecular Electronics, or in the area of quantum information circuits where these systems have been considered as spin entanglers \cite{Saraga} or quantum switches \cite{Suqing}.
\ack
E.C. acknowledges support from Proyecto DGAPA-UNAM IN110908 and from Programa de Intercambio Acad\'emico Internacional - UNAM. G.P. acknowledges
 support from the Spanish Research Ministry under grant MAT2008-02626 and from ITN
no. 234970 (EU).


\appendix

\section*{Appendix: Perturbative approach for the static case }

\setcounter{section}{1}

In this section we seek analytical expressions for the quasienergies using a perturbative approach for the static triple dot. We start from the Floquet equation corresponding to the Hamiltonian,
\begin{equation}
[\hat{H}''-i\hbar\partial /\partial
t]\,|\phi_i(t)\rangle=\epsilon_i\,|\phi_i(t)\rangle,
\label{eqflo}
\end{equation}
with $\hat{H}''=(\hat{H}_{0}+\hat{H}_{tun}+\hat{H}_{ac})$.

In this approach the tunneling $\hat{H}_{tun}$ is treated as a perturbation, i.e. $T \ll V_{ac}$ \cite{cref_plat}, using Rayleigh-Schr\"{o}dinger perturbation theory.
This procedure deals with the diagonalization of the perturbation matrix in the base of the unperturbed Floquet Hamiltonian ${\cal H}_{u}$ given by
${\cal H}_{u}=(\hat{H}_{0}+\hat{H}_{ac}-i\hbar\partial /\partial t)$.

We proceed to calculate the eigenvectors and eigenvalues of  ${\cal H}_{u}$ starting from the equation,
\begin{equation}
{\cal H}_{u}\,|\phi_i^u(t)\rangle=\epsilon_i^u\,|\phi_i^u(t)\rangle.
\label{eqflo_unp}
\end{equation}

The matrix elements of the unperturbed Hamiltonian in the dot basis ${\cal D}$
i.e. ${\cal H}_{u,qs}\equiv \langle q | {\cal H}_{u}|s\rangle $ ($q,s=l,c,r$) are given by,
\begin{eqnarray}
{\cal H}_{u,ll} &=&\frac{\varepsilon_b}{2}+\frac{V_{ac}}{2}\,cos\omega_{ac}t - i\hbar\frac \partial {\partial t}; \nonumber \\
{\cal H}_{u,cc} &=& -i\hbar\frac\partial {\partial t}; \nonumber \\
{\cal H}_{u,rr} &=&-\frac{\varepsilon_b}{2}-\frac{V_{ac}}{2}\,cos\omega_{ac}t - i\hbar\frac \partial {\partial t}; \nonumber \\
{\cal H}_{u,lc} &=& {\cal H}_{u,cl}=0;\nonumber \\
{\cal H}_{u,lr} &=& {\cal H}_{u,rl}=0;\nonumber \\
{\cal H}_{u,cr} &=& {\cal H}_{u,rc}=0,
\label{elemmatrizstatic}
\end{eqnarray}
The eigenvectors can be explicitly calculated from the matrix representation of equation (\ref{eqflo_unp}),
and to exemplify the procedure we consider the particular case for $\phi_l^u(t)$, which leads us to the equation,
\begin{equation}
\left[ \left( \frac{\varepsilon _b}2-\epsilon_l^u\right)
+(V_{ac}/2)\,cos\omega_{ac}t\right] \,\phi_l^u(t)=i\hbar \,\partial \phi_l^u(t)/\partial t,
\label{ecdifphi}
\end{equation}
which has the solution
\begin{equation}
\phi_l^u(t)=e^{-i\left( \varepsilon_b/2-\epsilon_l^u\right)
t/\hbar}\,e^{-i\xi \,sin\omega_{ac}t},
\label{solecdifphi}
\end{equation}
with $\xi =(V_{ac}/2\hbar \omega _{ac})$.
The corresponding eigenvector $\left| \phi_l^u \right\rangle $  reads,
\begin{equation}
\left| \phi_l^u \right\rangle =e^{-i\left( \varepsilon_b/2-\epsilon_l^u \right)
t/\hbar}\,e^{-i\xi \,sin\omega_{ac}t}\,\left| l\right\rangle
\label{phil}
\end{equation}
Similarly, we also obtain,
\begin{equation}
\left| \phi_r^u\right\rangle =e^{-i\left( -\varepsilon_b/2-\epsilon_r^u\right)
t/\hbar}\,e^{+i\xi \,sin\omega_{ac}t}\,\left| r\right\rangle,
\label{phir}
\end{equation}
and
\begin{equation}
\left| \phi_c^u\right\rangle=e^{i\epsilon_c^u\,t/\hbar}\,\left| c\right\rangle.
\label{phic}
\end{equation}
The eigenvalues are calculated using the periodicity of the Floquet states i.e. ($\phi_l^u(t)=\phi_l^u(t+2\pi /\omega _{ac})$) and are given by  $\epsilon_l^u=(\varepsilon_b/2+m^{(1)}\hbar \omega_{ac})$, $\epsilon_r^u=(-\varepsilon_b/2+m^{(2)} \hbar \omega_{ac})$ and $\epsilon_c^u=(0+m^{(3)}\hbar \omega_{ac})$, where $m^{(k)}=0,\pm1,\pm2,...$, for $k=1,2,3$.

We now proceed to calculate the matrix elements of the perturbation  $\hat{H}_{tun}$
as follows,
\begin{equation}
P_{q s}=\left\langle \left\langle \phi_q^u\left| \hat{H}_{tun}\right|
\phi_s^u \right\rangle \right\rangle =\frac 1\tau \int_0^\tau
dt\,\left\langle \phi_q^u\left| \hat{H}_{tun} \right| \phi_s^u\right\rangle,
\label{zambe}
\end{equation}
where $q,s=l,c,r$.
From the above equation it can be readily verified that $P_{ll}=P_{cc}=P_{rr}=P_{lr}=P_{rl}=0$, and the remaining non-zero matrix elements can be calculated by considering that an alignment of different Floquet quasienergy ladders occurs i.e. $\epsilon_c^u=\epsilon_r^u$, and $\epsilon_c^u=\epsilon_l^u$,  whenever the condition $\varepsilon_b=2s' \hbar \omega_{ac}$  ($s'=0,1,2,...$), is satisfied.
For the particular case of $P_{lc}$,
\begin{eqnarray}
P_{lc} &=&\left\langle \left\langle \phi_l^u\left| \hat{H}_{tun} \right|
\phi_c^u \right\rangle \right\rangle \\ \nonumber
&=&\frac 1\tau \int_0^\tau
dt\,\left\langle \phi_l^u\left| \hat{H}_{tun}\right| \phi_c^u\right\rangle,   \\ \nonumber
P_{lc} &=&T\,\frac 1\tau \int_0^\tau dt\,e^{+i\xi \,sin\omega
_{ac}t}\,e^{i\left( \varepsilon_b/2-\epsilon_l^u\right)
t/\hbar}\,e^{i\epsilon_c^u\,t/\hbar},
\label{api}
\end{eqnarray}
by using the identity
\begin{equation}
e^{\pm \,i\xi \,sin\omega _{ac}t}=\sum_{n=-\infty }^\infty J_n(\xi
)\,e^{\pm i\,n\,\omega _{ac}t},  \label{identbess}
\end{equation}
we obtain
\begin{equation}
P_{lc}=T\,\sum_{n=-\infty }^\infty J_n(\xi )\frac 1\tau \int_0^\tau
dt\,e^{i\omega_{ac}s't}\,e^{i\omega_{ac}nt},  \label{app3}
\end{equation}
which leads us to $P_{lc}=T\,J_{-s'}(\xi )=-T\,J_{s'}(\xi)$. Similarly, one can show that $P_{cl}=P_{cr}=P_{rc}=-T\,J_{s'}(\xi)$, thus the matrix elements of the perturbation are given by,
\begin{equation}
P=-T\,J_{s'}(\xi )\,\left[
\begin{array}{ccc}
0 & 1 & 0 \\
1 & 0 & 1 \\
0 & 1 & 0
\end{array}
\right],  \label{matrixp}
\end{equation}
and the diagonalization of equation (\ref{matrixp}) leads us to the set of eigenvalues $\{ 0, \pm \sqrt{2}\,T\,J_{s'}(\xi) \}$, where the result given by equation (\ref{perturb_stat}) follows.


\end{document}